\documentclass[english,aps,preprint]{revtex4}
\usepackage[T1]{fontenc}
\usepackage[latin9]{inputenc}
\usepackage{amsmath}
\usepackage{graphicx}

\makeatletter
\@ifundefined{textcolor}{}
{%
 \definecolor{BLACK}{gray}{0}
 \definecolor{WHITE}{gray}{1}
 \definecolor{RED}{rgb}{1,0,0}
 \definecolor{GREEN}{rgb}{0,1,0}
 \definecolor{BLUE}{rgb}{0,0,1}
 \definecolor{CYAN}{cmyk}{1,0,0,0}
 \definecolor{MAGENTA}{cmyk}{0,1,0,0}
 \definecolor{YELLOW}{cmyk}{0,0,1,0}
}

\makeatother

\usepackage{babel}
\begin{document}

\title{Quantum Complexity and Chaos in Young Black Holes}

\author{Alexander Y. Yosifov}
\email{alexanderyyosifov@gmail.com}

\affiliation{Department of Physics and Astronomy, Shumen University}

\author{Lachezar G. Filipov}
\email{lfilipov@mail.space.bas.bg}

\affiliation{Space Research and Technology Institute, Bulgarian Academy of Sciences}
\begin{abstract}
We argue the problem of calculating retention time scales in young
black holes is a problem of relative state complexity. In particular,
we suggest that Alice's ability to estimate the time scale for a perturbed
black hole to release the extra $n$ qubits comes down to her decoding
the Hilbert space of the Hawking radiation. We then demonstrate the
decoding task Alice faces is very difficult, and in order to calculate
the relative state complexity she would either need to act with an
exponentially complex unitary operator or apply an extremely fine-tuned
future precursor operator to the perturbed state in $SU(2^{K})$.
\end{abstract}
\maketitle

\subsection*{1. Introduction}

In recent years both quantum complexity and quantum information theory
have had a substantial presence in quantum black holes. 

In gauge/gravity duality a deep connection between holographic quantum
complexity and the horizon geometry has been proposed {[}1-4{]}. In
the framework of AdS/CFT, and assuming ER=EPR {[}5{]}, quantum complexity
has been argued to ensure infalling observers safe travels. The question
of whether or not high energy quanta are present behind the horizon
reduces to the question of whether Alice can decode a subfactor of
the Hilbert space of the Hawking radiation before the complexity bound
is saturated. So in the case of a two-sided AdS black hole, firewalls
are present if either Alice acts with a maximally complex (\emph{i.e.}
exponential in the entropy) unitary operator or if she waits for classical
recurrence time. 

Two important holographic duals have been proposed {[}4,5{]}, namely
the \textquotedbl{}complexity=action\textquotedbl{} and \textquotedbl{}complexity=volume\textquotedbl{}
conjectures. The former relates the complexity of a boundary CFT to
the action of the dual Wheeler-DeWitt patch in AdS, while the latter
relates the complexity of a holographic state to the volume of a maximally
extended spacelike hypersurface behind the horizon.

On the other hand, delocalization of quantum information between random
subsystems with respect to the Hilbert space factorization, and the
subsequent growth of entanglement is a central point for studying
the interior black hole region. Black holes with interior dynamics
described as a quantum circuit {[}6{]} have been proved to be fast
scramblers. In this framework they have been shown to scramble quantum
information in time logarithmic in the number of the degrees of freedom.
That is, the dynamics takes an initially localized perturbation and
makes it undetectable to an observer who fails to study a significant
part of the initial degrees of freedom. In turn, there is a growing
consensus that the scrambling time is the appropriate time scale associated
with release of quantum information.

In light of these advancements, we argue that calculating retention
time scales is actually a question of relative state complexity. We
claim that in such scenarios Alice cannot calculate the relative state
complexity before the complexity bound is saturated. Alice has two
options, she could either act with a maximally complex unitary operator
or act with a future precursor operator to the perturbed (late time)
state, and rely on extreme fine-tuning. Both options are shown to
be computationally unrealizable for evaporating black holes.

\subsection*{2. Black holes as random quantum circuits}

In the current Section we describe black holes as random quantum circuits
{[}6{]}. These are systems composed of $K$ degrees of freedom, and
have discrete time-step evolution $\varDelta\tau$, which is dictated
by a universal gate set of 2-local gates. A gate set is a collection
of gates (simple unitary transformations) which at each time-step
act on the qubit system. For simplicity we choose our gate set to
consist of 2-local gates, where each gate can act on no more than
2 qubits per time-step.

\subsection*{2.1 Fast scramblers}

Black holes have been proven to be the fastest scramblers in Nature
{[}6,10{]}. Scrambling, a form of strong thermalization, is a process
which stores information in highly nontrivial correlations between
subsystems. When chaotic thermal quantum systems of large number of
degrees of freedom undergo scrambling, their initial state, although
not lost, is very computationally costly to be recovered.

In this paper we assume that although the modes are scrambled they
are still localized in a certain way across the horizon, Fig. 1. However,
because of the strong thermalization, they remain \emph{indistinguishable}
from the rest of the black hole degrees of freedom as far as Alice
is concerned. Suppose Alice is outside the black hole, and throws
a few qubits inside. From her perspective, those extra modes will
be effectively diffused, \emph{i.e.} smeared across the horizon in
a scrambling time

\begin{equation}
t_{*}\sim\frac{\beta}{2\pi}\log N^{2}
\end{equation}
where $\beta$ is the inverse temperature, and $N$ is the number
of degrees of freedom. 

As a result, a scrambling time after perturbing the black hole, Alice
will not be able to distinguish those extra qubits. This statement
is similar to the upper chaos bound for general thermal systems in
the large $N$ limit {[}22{]}. In particular, the large $N$ factor
is what initially keeps the commutators small. However, for $t>t_{*}$,
scrambling yields rapid commutator growth, and so the distance between
the initial and perturbed states in complexity geometry increases
non-trivially, (12,13). 
\begin{figure}
\includegraphics[scale=0.6]{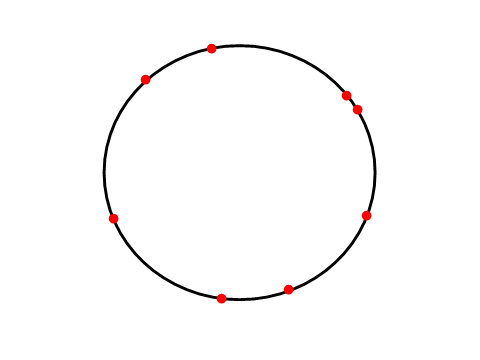}\caption{Depiction of a black hole as an $(K+n)$-qubit system. Imagine the
region inside the circle is the black hole interior, and the one outside
of the circle is the exterior region. The red dots are the scrambled
extra $n$ qubits embedded into the horizon. The present figure was
inspired by Fig. 10 from {[}2{]}.}
\end{figure}

In strongly coupled thermal quantum systems chaos and scrambling are
intimately related, which is why (1) is of particular interest to
both quantum cloning, and retention time scales, \emph{i.e.} the minimum
time for information to begin leaking via Hawking radiation. 

Imagine Bob crosses the horizon carrying a qubit, and Alice hovers
outside the black hole. It was shown {[}6{]} that by the time Alice
recovers the perturbed qubit by collecting the outgoing modes, and
enters the black hole, Bob will have already hit the singularity.
So retention time scale of order the logarithm of the entropy (2)
is just enough to save black hole complementarity from paradoxes

\begin{equation}
t_{ret}\geq\frac{\beta}{2\pi}\log N^{2}
\end{equation}
Thus quantum cloning cannot be verified given the above bound is respected.

Recent studies of quantum information and quantum gravity {[}6,7,8,12,13,14,15,16,17{]}
support the scrambling time as the appropriate time scale at which
black holes begin releasing information via Hawking radiation. Note
that in such generic early evaporation scenarios where quantum information
begins leaking out of order the scrambling time, not every Hawking
particle carries information as this would make the retention time
scale $t_{ret}\sim\log r_{S}$ , which would violate the no-cloning
bound (2).

\subsection*{2.2 Qubit description of black holes}

A quantum circuit is composed of gates, and describes the evolution
of a quantum system of qubits. The gates may be defined to act on
an arbitrary number of qubits, and to couple any given pair of them.
The gates may act in succession or in parallel, where series-type
quantum circuits are not good scramblers.

Here, we present a random quantum circuit with time-dependent Hamiltonian
which has been proved to scramble in time logarithmic in the entropy
{[}11{]}.

Consider a $K$-qubit analog of a Schwarzschild black hole in 3+1
dimensions, where 

\begin{equation}
K\sim S_{BH}\sim\frac{A}{4G_{N}}
\end{equation}
here $A$ is horizon area, and $G_{N}$ is Newton's constant. 

Let the $K$ qubits be in some initial pure state of the form

\begin{equation}
\left|\psi\right\rangle =\sum_{i}\alpha_{i}\left|i\right\rangle 
\end{equation}
where $\alpha_{i}$ is the amplitude, and $\left|i\right\rangle $
is the Hilbert space basis. The state lives in a Hilbert space of
$2^{K}$ dimensions.

In this framework the Hamiltonian is given as {[}11{]}

\begin{equation}
H_{i}=\sum_{l<m}\sum_{\alpha_{l},\alpha_{m}=0}^{3}\sigma_{l}^{\alpha_{l}}\otimes\sigma_{m}^{\alpha_{m}}\varDelta B_{i,l,m,\alpha_{l},\alpha_{m}}
\end{equation}
where $\varDelta B_{i,l,m,\alpha_{l},\alpha_{m}}$ denote the independent
Gaussians, $\sigma_{i}$ are Pauli matrices, and the eigenenergies
live in a $2^{K}$ dimensional state space.

Thus the evolution between 2 successive time-steps is 

\begin{equation}
e^{-iH_{i}\Delta\tau}e^{-iH_{i+1}\Delta\tau}
\end{equation}
Here, there is an inverse relation between the time-step and the strength
of the interactions. It was thus demonstrated by Hayden et. al. in
{[}11{]} that the time required to scramble an arbitrary number of
degrees of freedom scales like $\log k$.

The evolution of the $K$-qubit system is controlled by a random quantum
circuit, composed of a universal gate set of 2-local gates, where
we assume the gate set approximates Hamiltonian time evolution

\begin{equation}
U=\left\{ g_{i}\right\} 
\end{equation}
Suppose $k$-local gates with $k>2$ are strictly penalized. In addition
to the $k$-local restriction, we assume the gates have non-zero couplings
only with respect to the nearest qubits, similar to ordinary lattice
Hamiltonians. Of course, in principle, nothing demands this particular
locality constraint, and we could have easily allowed any arbitrary
pair of qubits to couple.

The evolution is divided into time steps $\varDelta\tau$, where at
each \emph{time-step} a random gate set is chosen. The choice is random
because at each time-step the gate set is picked via a time-dependent
Hamiltonian governed by a stochastic probability distribution. Furthermore,
the random choice has to also determine which $k$ qubits the gate
set will act on. Note that the random quantum circuit that we use
is bounded from above by $K/2$ gates which are allowed to act in
parallel every time-step. 

We suggest a natural time scale to associate with the time intervals
between successive time steps would be the Schwarzschild radius $\varDelta\tau\sim r_{S}$
(i.e. light crossing time). One does not have to look further than
elementary black hole mechanics to see why this is the case. For instance,
in a freely evaporating black hole, $r_{S}$ is adiabatically decreasing.
Consequently, the time intervals between successive time steps are
shorter, and thus black hole evaporates at a faster rate. This fits
well with classical black hole thermodynamics

\begin{equation}
r_{S}\sim T^{-1}\sim\beta
\end{equation}
where $\beta$ denotes the inverse temperature.

Since this random quantum circuit scrambles information logarithmically,
throughout the paper we consider it to be an effective analog of general
early evaporation models.

\subsection*{2.3 Relative state complexity}

In this Section we argue that calculating retention time scales for
systems controlled by random quantum circuits is a question of relative
state complexity. We show that calculating relative state complexity
is extremely difficult, and in order for Alice to carry out the computation
she would need to either act with an exponentially complex unitary
operator or apply a future precursor operator to the perturbed state,
and rely on extreme fine-tuning.

We can simply define circuit (gate) complexity as the minimum number
of gates required to implement a particular state. The evolution of
a quantum state via time-dependent Hamiltonian resembles the motion
of a non-relativistic particle through the homogeneous $SU(2^{K})$
space {[}18{]}. That is, the particle defines a trajectory between
a pair of points, where one point corresponds to some initial state
$\left|\psi\right\rangle $, while the second point corresponds to
a perturbed state $\left|\psi'\right\rangle $, where $\left|\psi\right\rangle ,\left|\psi'\right\rangle $
$\in SU(2^{K})$. The particle thus moves on a $2^{K}$ dimensional
state space. Without loss of generality the state evolution can be
straightforwardly given by the Schrodinger equation

\begin{equation}
i\frac{\partial\left|\psi\right\rangle }{\partial t}=H\left|\psi\right\rangle 
\end{equation}
Given two states are different to begin with, their relative state
complexity naturally increases with time.

A compelling argument was made in {[}18{]} that the naive way of defining
the distance between two states does not capture the whole story.
The classical Fubini metric bounds the state distance as $d\in[0,\pi/2]$.
Obviously, the upper bound can be easily saturated and we need a different
measure to quantify the relative state complexity between two states\emph{.
}Due to the exponential upper bound of complexity, $C_{max}=e^{K}$,
quantifying relative state complexity necessitates the use of \emph{complexity
metric}. That is the notion of distance between states on the non-standard
$SU(2^{K})$, see Refs. {[}19,20{]}. Intuitively, the farther apart
two states are on $SU(2^{K})$, the higher their relative state complexity
is. Geometrically, we can think of relative state complexity as \emph{the
minimum geodesic length in $SU(2^{K})$ which connects two states.}

In light of the proposed random quantum circuit, and following the
definition of circuit complexity, we define relative state complexity
as \emph{the minimum number of time steps required to go from one
quantum state to another.} 

Keep in mind that every time step a random set of 2-local gates is
chosen following a time-dependent Hamiltonian controlled by a stochastic
probability distribution. So essentially, using Nielsen's approach
{[}19{]}, we are interested in assigning a notion of distance (geodesic
length) to the gate sets. Here the minimum length increase in complexity
geometry sets a lower bound for the minimum complexity growth, which
corresponds to acting with a single 2-local gate.

More precisely, suppose Alice perturbs the $K$-qubit system immediately
after its formation by $n$ qubits, where $n\ll K$, Fig. (1). We
ask: what is the relative state complexity of $\left|\psi\right\rangle $
and $\left|\psi'\right\rangle $? In other words, what is the minimum
number of time steps $N(\varDelta\tau)$ in which Alice could time-reverse
the perturbation

\begin{equation}
\left|\psi\right\rangle =U_{1}U_{2}U_{3}...U_{N(\Delta\tau)}\left|\psi'\right\rangle 
\end{equation}
where $\left|\psi\right\rangle \in$ $K$-qubit system, $\left|\psi'\right\rangle \in$
$(K+n)$-qubit system.

Let's now turn to the main objective of this paper which is to address
the question:

$\vphantom{}$

\emph{In a young black hole, can Alice calculate the relative state
complexity of $\left|\psi\right\rangle $ and $\left|\psi'\right\rangle $
in time less than $2^{K}$? }

$\vphantom{}$

In our case calculating the relative state complexity means counting
the number of time steps in which the extra $n$ qubits are radiated
to infinity.

Our claim is that in implementing $U$, Alice cannot beat $2^{K}$
because of the causal structure of the black hole spacetime. Alice
does not have access to all the relevant degrees of freedom which
dramatically increases the computational complexity. Therefore, the
inability of implementing $U$ faster than $2^{K}$ not only renders
the computation unrealizable for astrophysical black holes, but also
takes away Alice's predictive power.

We will now look at the two ways Alice could hope to estimate (9).
Namely, she could either apply gate sets to the radiated qubits, or
act with an extremely fine-tuned precursor.

\subsubsection*{2.3.1 How fast can Alice calculate the relative state complexity?}

In this subsection we consider the Harlow-Hayden reasoning {[}16{]}
but for the case of a young black hole. Recall that in {[}16{]} Harlow
and Hayden argued that AMPS' conjectured violation of the equivalence
principle after Page time is computationally unrealizable for black
holes formed by sudden collapse since it requires complicated measurements
on the emitted Hawking cloud. We now study the limit of the proposed
$2^{k+m+r}$ complexity bound, and demonstrate that it is strong enough
to hold even for the case where (i) entanglement entropy is still
low, and (ii) $\mathcal{H}_{R}\ll\mathcal{H}_{BH}$.

Here we employ a standard Hilbert space decomposition where $k$ are
the black hole qubits in $\mathcal{H}_{BH}$, $m$ are the qubits
of the black hole atmosphere in $\mathcal{H}_{B}$, and $r$ are the
emitted qubits in $\mathcal{H}_{R}$, whose dimensionality grows as
the black hole evaporates. We assume Alice can only manipulate the
$r$ qubits, and that all outside observers must agree on $\mathcal{H}_{B}\otimes\mathcal{H}_{R}$.
For her this is essentially a decoding problem, where she acts with
the unitary transformation $U$ to $\mathcal{H}_{R}$. Alice's goal
is to decode $\mathcal{H}_{R}$ in search for the extra $n$ qubits,
and count the number of time steps in which they were radiated away.

To demonstrate more clearly the robustness of the $2^{k+m+r}$ complexity
bound, suppose we violate the fast scrambling conjecture {[}11{]}.
The violation is in the sense that Alice can recognize the perturbed
$n$ qubits easier, and doesn't need to decode a significant part
of all the system's degrees of freedom. Even in this case, however,
we argue there is an overwhelming probability Alice \emph{cannot}
beat $2^{k+m+r}$. 

Since the scrambling time is the shortest time-scale compatible with
the no-cloning bound (2), one might naively expect Alice can time-reverse
the perturbation in time comparable to the scrambling time. Considering
the exponentially high upper bound of complexity, however, we can
easily see that this is not the case {[}2{]}. Even though the scrambling
time is negligible compared to the time-scale associated with reaching
maximum complexity, in a scrambling time the complexity of the system
scales as

\begin{equation}
C_{*}=S\log S
\end{equation}
Although nowhere near the upper bound of $C_{max}=e^{K}$, the scrambling
complexity is high enough to make the computation extremely difficult.
From a geometry perspective, by the scrambling time, due to quantum
chaos, the trajectories of the 2 points on $SU(2^{K})$ diverge exponentially,
Fig. 2.
\begin{figure}
\includegraphics[scale=0.53]{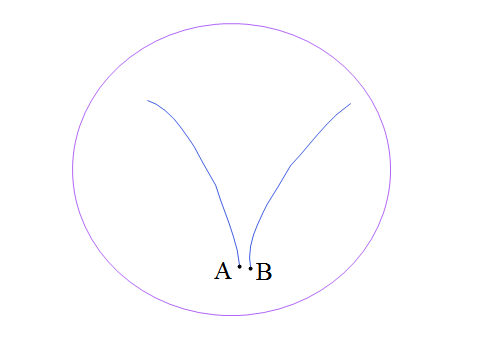}

\caption{A pair of points and their trajectories on complexity geometry $SU(2^{K})$.
They are initially arbitrarily close, i.e. low relative state complexity.
At the scrambling time, however, their trajectories diverge, and the
distance between them grows exponentially. }
\end{figure}
 Despite being initially arbitrarily close, given they are separated
to begin with, in just a scrambling time the distance (i.e. relative
state complexity) between them grows exponentially. 

Let's further illustrate the point with the use of an out-of-time-order
correlator (OTOC) $C(t)$. OTOCs are used for measuring quantum chaos
{[}24,25{]}. In particular, OTOCs describe how initially commuting
operators develop into non-commuting ones. Suppose $A(0)$ and $B(t)$
are simple Hermitian operators, where $B$ is just a time-evolved
$A$. Initially, for $t\ll t_{*}$ the correlator is approximately
constant. Then at the scrambling time, due to quantum chaos {[}22{]}

\begin{equation}
C(t)=-\left\langle (B(t),A(0))^{2}\right\rangle _{\beta}
\end{equation}
After the scrambling time, regardless of $A$ and $B$, the correlator
takes the form 
\begin{equation}
C(t)=2\left\langle BB\right\rangle \left\langle AA\right\rangle 
\end{equation}
This exponential decay is associated with the initial rapid growth
of the commutator, which becomes highly non-trivial at the scrambling
time. Note that for small $t$, the large number of black hole degrees
of freedom suppress the commutator. Therefore, the scrambling time
is enough to make the operators very complicated, and thus the distinguishablity
between them non-trivial. So what can Alice do?

The obvious thing to do would be to brute-force. In this case Alice
will have to first artificially group the $r$ qubits in $\mathcal{H}_{R}$
into different sets, and then apply a complex unitary transformation
to those sets in search for the extra $n$ qubits. The difficulty
here is to have a unitary transformation which acts on a particular
set of qubits {[}16{]}. Unlike the Harlow-Hayden argument where Alice
tries to decode a subfactor of $\mathcal{H}_{R}$ in order to verify
entanglement with $\mathcal{H}_{B}$, here the decoding task is especially
complicated given Alice will have to engineer multiple such unitary
transformations because of the monotonically growing dimensionality
of $\mathcal{H}_{R}$. It seems that even with the assumption we made
that Alice need not probe a significant part of all the initial degrees
of freedom to recognize the extra qubits, the computation remains
extremely non-trivial.

Of course, Alice could try to use some clever tricks to carry out
the computation faster than $2^{k+m+r}$. For instance, she could
try to impose some special structure to the unitary transformation,
and in particular, on how it evolves with time. She could engineer
$U$ to allow specific sequences of gate sets to act every time-step
on preferred qubit sets. However, such modifications can only account
for very small changes which are not enough to speed up the computation
to reasonable time scales.

Another possibility is for Alice to manipulate the adiabatically growing
number of $r$ qubits and make them interact in a preferred manner.
By making use of the smaller dimensionality of $\mathcal{H}_{R}$,
she could form sets of qubits, establish certain connection between
them, and choose which sets to interact every time step. However,
despite $r\ll k$ engineering such a connection between the $r$ qubits
would obviously require introducing additional degrees of freedom
which would scale like an exponential in $r$. Thus the computation
becomes very complex even for relatively small $r$. In fact, by trying
to fine-tune the qubits in this way, Alice makes her decoding task
harder.

Clearly, even in the case of a young black hole, and making the unphysical
assumption that Alice need not decode a large part of the entropy,
the computation remains very hard. Even though there is nothing, in
principle, that prevents the first $n$ emitted qubits to be the perturbed
one, this is exponentially unlikely. So even with weak violations
of the fast scrambling conjecture the $2^{k+m+r}$ bound holds with
an overwhelming probability.

What about black holes which have evaporated more than half of their
initial degrees of freedom? One could hope to speed up the computation
by letting the black hole evaporate passed its Page time, and apply
certain gate sets on $\mathcal{H}_{R}$. For example, once $\mathcal{H}_{R}>\mathcal{H}_{BH}$,
ancillary qubits could be introduced and entangled with subfactors
of $\mathcal{H}_{R}$. Then one could apply gates to those subfactors
in effort to implement $U$ faster than $2^{k+m+r}$. Unfortunately,
as long as the extra qubits scale like $r$, the computation does
not get any faster. 

Therefore, unless Alice finds a way to calculate the relative state
complexity in a way which does not involve exponential number of gates,
the $2^{k+m+r}$ time-scale remains solid.

\subsubsection*{2.3.2 Precursors and extreme fine-tuning}

Here we examine Alice's second attempt of calculating the relative
state complexity which now involves applying a future precursor operator
to the late-time state. For simplicity, we study the case using a
generic time-independent Hamiltonian but expect the main conclusions
to hold for the time-dependent ones, too.

Alice's task here is to adjust the late-time state and time reverse
it. Effectively, this means running the chaotic black hole dynamics
backwards. We show that this process of time-reversing (10) by applying
a future precursor operator to the perturbed state is notoriously
difficult, and Alice still cannot beat $2^{k+m+r}$. The particular
argument should be considered in the context of complexity geometry
on $SU(2^{K})$.

The laws of physics are time-reversible, so any state perturbation
that we introduce could be reversed. Naturally, however, for $t>t_{*}$
complexity tends to increase linearly in $K$ until it saturates the
bound of $C_{max}=e^{K}$. Therefore, after the scrambling time we
expect linear increase of the relative state complexity between $\left|\psi\right\rangle $
and $\left|\psi'\right\rangle $ as $t$ grows. Geometrically, this
corresponds to a linear growth of the minimum geodesic length connecting
the two states in $SU(2^{K})$. 

Let's analyze the same example of Alice perturbing the $K$-qubit
system immediately after its formation with $n$ qubits. As we already
saw, whatever Alice does, she cannot carry out the computation faster
than $2^{k+m+r}$. Determined to calculate the relative state complexity
before the complexity bound is saturated, however, imagine she now
acts with a specific operator, namely a future precursor operator
{[}21{]}.

A future precursor operator $P_{p}^{+}$ is a highly non-local operator
which when applied at a certain time simulates acting with a local
operator $P$ at an earlier time

\begin{equation}
P_{p}^{+}=U(t)PU^{\dagger}(t)
\end{equation}
where $U(t)=e^{-iHt}$.

Generally, calculating a precursor operator for $\Delta t\geq t_{*}$
is extremely difficult as one has to keep track of all the interactions
of the degrees of freedom of the system. The computational costs grow
immensely in cases involving black holes because they not only have
a large number of degrees of freedom but also saturate the chaos bound
{[}22{]}. For the first scrambling time after perturbing a black hole,
the complexity growth is governed by the Lyapunov exponent, and hence
grows exponentially {[}23{]}. Black holes are the fastest scramblers
in Nature, and due to their chaotic dynamics, only a scrambling time
after the perturbation, all of the degrees of freedom ($K\sim10^{77}$
for a solar mass black hole) will have indirectly interacted. Evidently,
the precursor operator quickly becomes extremely difficult to calculate.
Whatever Alice does to implement the precursor, she must time-reverse
all of the interactions between the degrees of freedom of the black
hole which requires an extreme degree of fine-tuning. Regardless of
the exponential complexity, however, individual interactions remain
well defined.

In our case acting with the precursor operator takes the general form 

\begin{equation}
\left|\psi\right\rangle =e^{-iHt}Pe^{iHt}\left|\psi'\right\rangle 
\end{equation}
Similar to the evolution of a quantum state via time-dependent Hamiltonian,
the action of a future precursor resembles a backward motion of a
particle through complexity geometry. 

The high complexity of (14) corresponds to the complexity associated
with constructing a thermofield-double state using only $t<0$ degrees
of freedom, see Ref. {[}21{]}. In both cases, due to the large number
of degrees of freedom an extreme fine-tuning is required. Even a mistake
of order a single qubit will accumulate, and result in a completely
different end-state. The system only becomes more sensitive to errors
as the time separation increases. Therefore, unlike regular unitary
operators which need not always be complex, precursors are typically
extremely complex and unstable to perturbations (the butterfly effect)
whenever the time separation is at least of order the scrambling time. 

Expanding (15) for $\Delta t\sim t_{*}$ yields

\begin{equation}
\left|\psi\right\rangle =e^{-iH(t_{*}-t_{i})}Pe^{iH(t_{*}-t_{i})}\left|\psi'\right\rangle 
\end{equation}
where $t_{i}$ is the initial time when the $K$-qubit system was
perturbed.

Notice we have restricted our analysis not to include cases when either
$\Delta t\ll t_{*}$ or $\Delta t\gg t_{*}$. The former case was
discussed in Ref. {[}23{]}, where it was argued that before the scrambling
time the distance in complexity geometry between the initial and perturbed
states remains approximately constant. Initially, for $t<t_{*}$,
the large $N$ terms keep the commutators relatively small. So scrambling
is what drives the rapid distance growth in $SU(2^{K})$. On the other
hand, the latter case is unnecessary since, as we showed, the computation
becomes unmanageable in just $t_{*}$.

Therefore, due to the chaotic black hole dynamics, and the great deal
of fine-tuning required, the probability of Alice implementing the
precursor (and thus calculating the relative state complexity) without
making a mistake of even a single qubit is exponentially small.

In conclusion, we can see that due to the causal structure of the
black hole geometry, and the chaotic dynamics there is nothing Alice
can do that would allow her to calculate the relative state complexity
faster than $2^{k+m+r}$. This exponential time scale, however, is
only applicable for AdS black holes since astrophysical black hole
evaporate much before the complexity bound is saturated. So in the
case of black hole formed by a sudden collapse there are two very
general scenarios, associated with minimum and maximum retention time
scales, $t_{min}$ and $t_{max}$, respectively. Obviously, the fastest
retention time possible which obeys the no-cloning bound (2) is

\begin{equation}
t_{min}\sim\mathcal{O}\left(\frac{\beta}{2\pi}\log N^{2}\right)
\end{equation}
up to some constant. 

This is similar to the Hayden-Preskill result {[}6{]} concerning the
mirror-like dynamics of an old black hole. On the other hand, the
longest retention time for astrophysical black holes would be of order
the evaporation time $t_{ev}$ 

\begin{equation}
t_{max}\sim\mathcal{O}(M^{3})
\end{equation}
Usually, such retention time scales are associated with remnants which
have been seriously questioned due to the apparent violation of the
Bekenstein entropy bound. 

\subsection*{3. Conclusions}

The goal of this paper was to argued that calculating retention time
scales is a decoding task, and a problem of relative state complexity.
Our claim was that Alice cannot calculate the relative state complexity
between an initial and perturbed states before the complexity bound
is saturated. We considered a quantum system of $K$ qubits whose
interactions are dictated by a random quantum circuit, and assumed
the gate sets approximate a Hamiltonian evolution. In this framework,
at every time-step the quantum circuit implements a random set of
2-local gates according to a time-dependent Hamiltonian with stochastic
probability distribution. In this setting we perturbed the $K$-qubit
system with $n$ qubits, and assumed that (i) $\mathcal{H}_{R}\ll\mathcal{H}_{BH}$,
(ii) the black hole begins releasing information a scrambling time
after its formation, and (iii) nothing in principle prevents the first
$n$ emitted qubits to be the perturbed ones.

We examined several techniques Alice could use to decode $\mathcal{H}_{R}$,
and showed she cannot beat $2^{k+m+r}$. We demonstrated there is
an overwhelming probability Alice cannot decode $\mathcal{H}_{R}$
in time less than $2^{k+m+r}$, unless she acts with an exponentially
complex unitary operator or apply an extremely fine-tuned future precursor
operator to the perturbed state in $SU(2^{K})$, which renders the
computation unrealizable for evaporating black holes. In summary,
we made the case that the $2^{k+m+r}$ bound proposed by Harlow and
Hayden {[}16{]} holds strong even for young black holes.

\end{document}